# Work-Function Mapping Across a-In$_2$Se$_3$ → α-In$_2$Se$_3$ → γ-InSe in RF-Sputtered Thin Films

Marius O. Eji[1], Md. Sakauat Hasan Sakib[1], & Joseph P. Corbett[1]

[1]Department of Physics, Miami University, Oxford, OH 45056

**Abstract**

Indium selenide is a phase-change chalcogenide whose polymorphism enables a variety of physical properties to be tuned. Here we directly quantify the evolution of the surface work function across the amorphous-to-crystalline transition in RF-sputtered In$_2$Se$_3$ thin films grown on c-plane Al$_2$O$_3$ (001). By varying deposition temperature (100-500 °C) and film thickness, we establish processing windows for a- In$_2$Se$_3$, α-In$_2$Se$_3$, and γ-InSe, and correlate structure with electronic and morphological properties. X-ray diffraction shows films deposited at 100–200 °C are amorphous, 300–400 °C yields α-In$_2$Se$_3$, whereas 500 °C yields γ-InSe. Williamson–Hall analysis reveals a non-monotonic crystallite-size dependence on thickness and, with temperature, an onset before crystallites become resolvable by XRD. Kelvin probe force microscopy (KPFM) maps the surface potential and yields spatially averaged work functions spanning 5.26–6.64 eV across the amorphous–crystalline transformation; pronounced intra-film heterogeneity is observed, with select α-phase and γ-InSe grains exhibiting work functions exceeding the local mean. Topographical distinctions are found between phases with hexagonally faceted grains in the crystalline state, whereas homogeneous nano-mounds are found in amorphous films. Analysis of Tauc plots revealed optical bandgaps in the range of 2.50–1.55 eV across the observed phases. X-ray fluorescence (XRF) measurements further indicated that the indium-to-selenium concentration ratio varied between $0.70 \pm 0.1$ and $1.01 \pm 0.1$ as the deposition temperature increased from 100 to 500 °C. These measurements provide direct, spatially resolved quantification of work-function evolution through the phase change, supplying parameters essential for contact engineering and device integration of In$_2$Se$_3$ and InSe.

## I. INTRODUCTION

In$_2$Se$_3$ is an important group III-VI semiconductor material that has gained tremendous attention owing to its unique optical, mechanical, and electrical properties.[1] Its electrical properties depend on the crystal structure.[2] It is the most attractive 2D material due to its excellent physical properties making it favorable material for solar energy conversion, photodetectors, electronics, and optoelectronics applications.[3–5] In$_2$Se$_3$ undergoes reversible phase-change (amorphous ↔ crystalline [6,7]), which makes it potential material for phase change memory applications.[8] It is polymorphic in nature with intricate crystalline structures[3], and exists in various phases such as α, β, γ, δ, and κ phases.[8–10] The presence of cation vacancies play a significant role in determining these crystal structures of In$_2$Se$_3$.[11] Among its distinct phases, the α-In$_2$Se$_3$ phase is particularly intriguing, due to its potential applications in many fields such as optoelectronics [12,13], photovoltaics [14,15], and nonvolatile memory [16,17]. The α-In$_2$Se$_3$ phase is stable at room temperature with a direct bandgap energy of 1.3 eV[9] with the space group R3m (160).[11,18,19]

Indium Selenide (InSe), on the other hand, is also layered material, widely studied for its diverse crystal phases and promising applications in optoelectronics.[20–22] It exists in multiple structural phases, most notably γ-InSe, β-InSe, and ε-InSe by their stacking sequence and electronic properties.[20,23] Although they differ in stacking sequences, all three polytypes share an identical in-plane configuration, a hexagonal monolayer structure with a lattice constant of approximately 4.00 Å.[23] Along the out-of-plane axis, γ-InSe adopts an ABC stacking sequence, forming a 3R rhombohedral structure with a lattice constant of approximately 25.32 Å, while β-InSe and ε-InSe exhibit AB stacking typical of a 2H hexagonal structure, with a reduced lattice constant of about 16.64 Å.[23] InSe, despite its desirable electronic and optical properties, remains a relatively unstudied 2D semiconductor, primarily as high-quality samples have been obtained via mechanical exfoliation rather than large area thin-film growth. Although, recent work has shown optical grade large area deposition by pulsed laser deposition [23].

Both α-$In_2Se_3$ and γ-InSe crystallize in the rhombohedral space group R3m[24–26] and exhibit a van der Waals layered structure, characterized by strong in-plane bonding and weak interlayer interaction. This shared symmetrical and layered architecture enables low-temperature epitaxial growth and phase-specific transitions, which are critical for constructing vertical heterostructures with tunable properties.[26] The 3D structure of α-$In_2Se_3$ and γ-InSe (see, Figure 1) shows that γ-InSe has a layered hexagonal structure with four atomic planes per layer (Se-In-In-Se), while α-$In_2Se_3$ has a more complex quintuple-layered structure (Se-In-Se-In-Se), which lacks a center of symmetry.

Beyond their polymorphism, $In_2Se_3$ exhibits a reversible amorphous ↔ crystalline phase change [6,7] positioning it as a candidate for phase-change memory [8] and tunable optoelectronic platforms. [3–5] Realizing these technologies requires process–structure–property maps that connect deposition parameters to phase formation and, critically, to electronic figures of merit at device-relevant interfaces. Prior studies have explored α-phase stabilization via temperature gradients [10], and have successfully fabricated amorphous films using RF magnetron sputtering[27] and thermal evaporation.[28] Additionally, crystalline InSe has been synthesized via pulsed laser deposition (PLD)[23], molecular beam epitaxy (MBE)[29], and chemical vapor deposition (CVD)[30], but a direct spatially resolved quantification of how the surface work function evolves through the amorphous to crystalline α-$In_2Se_3$, and γ-InSe transition has remained limited.

Here we address this gap using RF magnetron sputtering to grow In-Se thin films on c-plane $Al_2O_3$(0001) while systematically varying deposition temperature and thickness. X-ray diffraction (XRD) establishes processing windows for a-$In_2Se_3$ to α-$In_2Se_3$ to γ-InSe and, via Williamson–Hall analysis, reveals a non-monotonic crystallite-size evolution with thickness together with a temperature onset for crystallinity. Atomic force microscopy (AFM) distinguishes homogeneous mounded topographies in amorphous films from hexagonally faceted grains in α-$In_2Se_3$ and γ-InSe. Tauc plots generated from Ultraviolet–visible spectroscopy (UV–vis) analysis allowed us to quantify concomitant changes in the optical gap across the transition, while the elemental composition was investigated using X-ray fluorescence (XRF) spectroscopy. Most importantly, Kelvin probe force microscopy (KPFM) maps the contact potential difference to yield spatially averaged surface work functions spanning ~5.26–6.64 eV from amorphous to crystalline films, with pronounced intra-film heterogeneity wherein select α-phase and γ-InSe grains exceed the local mean. To our knowledge, these results provide the first direct, spatially resolved measurement of the work-function evolution through the a-$In_2Se_3$→α-$In_2Se_3$→γ-InSe transition.

By linking controllable growth conditions to phase selectivity and to a device-critical interfacial parameter (work function), this study delivers actionable guidance for contact engineering, Schottky-barrier tuning, and integration of $In_2Se_3$ and InSe in phase-change and optoelectronic architectures.

## II. EXPERIMENTAL

### A. *Thin Film Growth*

The samples were deposited using the radio frequency (RF) magnetron sputtering. Under controlled conditions, we utilized high-purity 4N indium selenide ($In_2Se_3$) targets which were produced in-house via a 40-Ton isostatic press. Materials were weighted to a 0.0001 g precision to a 2:3 stoichiometry, and pressed under vacuum ($1\times10^{-2}$ Torr) at 250 C⁰ for 1 week. Single-sided and double-sided polished C-plane $Al_2O_3$(001) substrates were used for the deposition. To ensure maximum adhesion and film quality, the substrates were thoroughly cleaned with ethanol to remove impurities on the surface. The sputtering process was conducted in two ways: in the first case, with double-sided polished C-plane $Al_2O_3$ (001) substrates, the temperature was varied (100 °C, 200 °C, 300 °C, 400 °C, and 500 °C) at constant thickness (200 nm). In the second case, with single-sided polished C-plane $Al_2O_3$ (001) substrates, we kept the temperature constant (500 °C) and varied the thicknesses between 47 nm and 564 nm. The chamber was evacuated to a base pressure of $1.0\times10^{-8}$ Torr before introducing 6 mTorr argon gas to generate the plasma. The RF power source was kept at 5 W throughout the deposition with 0W power reflected.

### B. *Structural Characterization*

Phase identification was performed by Bruker D8 Advanced XRD using Cu Kα radiation ($\lambda$ = 1.5406 Å). Coupled symmetric θ–2θ scans were collected over 2θ = 4–120° with a step size of 0.02° and a count time of 1 s per step. Interplanar spacings were obtained from Bragg's law, and lattice parameters **a** and **c** were refined using standard hexagonal relationships between miller indices and d-spacing. Crystallite sizes were estimated by Williamson–Hall analysis from the integral breadth of 00l reflections.[31–34]

### C. *Elemental Composition*

The elemental composition of the films was analyzed using X-ray fluorescence (XRF) spectroscopy. Instrument stability and accuracy were verified through measurements on pure indium and selenium standards, as well as blank c-plane substrates. Spectra were obtained over the energy range of 0-60 keV with a counting time of 120 seconds per sample. Peak fitting and background subtraction were performed using *Fityk* software to improve precision and isolate elemental contributions.[35] Quantitative analysis was carried out, and stoichiometry was determined using the relative intensities of the detected indium and selenium peaks from our samples, and the sensitivity of pure indium and selenium.

### D. *Optical Characterization*

The optical properties of the synthesized $In_2Se_3$ samples were characterized using Ultraviolet-visible (UV-vis) spectroscopy. Absorption spectra were collected over the wavelength range of 325–1100 nm at room temperature. Each sample was placed in a custom 5 mm path length cuvette. Prior to measurement, baseline correction was performed using a blank c-plane substrates as reference to eliminate background absorption and noise. The resulting absorbance spectra were analyzed to estimate the optical bandgap energy using the Tauc plot method.[36,37]

### E. *Scanned Probe Measurements*

A commercial high-resolution (15 pm noise floor) multimodal atomic force microscope by AFM Workshop, with accompanying KPFM mode was utilized to investigate the morphology and work function. KPFM was used to map the contact potential difference (CPD) through an amplitude modulation second pass method.[38] The measurement was conducted under ambient conditions using a conductive platinum-coated cantilever with a nominal force constant of 3 N/m and a resonance frequency of 75 kHz. During the KPFM scan, the lift height was maintained at 20 nm to minimize topographical interference. An AC voltage of 1 V at 64.4 kHz was applied between the tip and the sample surface, allowing for the detection of contact potential difference (CPD) through electrostatic force modulation. The resulting surface potential maps were processed and analyzed using *Gwyddion* software.[39]

To calculate the work function of $In_2Se_3$ and InSe sample, we first performed a series of tip calibration scans on a gold-coated reference sample with an externally applied DC offset voltages (1.5–4.5 V), and recorded the corresponding contact potential differences (CPD).[40] By applying the known work function of gold[41] and the measured CPD from the reference sample, we calculated the tip's work function. This value was then used to compute the work functions of both amorphous, $\alpha$-$In_2Se_3$, and $\gamma$-InSe based on the CPD values obtained from their respective KPFM measurements.

## III. RESULTS AND DISCUSSION

### A. XRD and XRF Analysis

*Figure 2* **(a)** is the deposition on the double-sided polished c-plane $Al_2O_3$ (001) substrates at constant thickness 200 nm. Crystalline phases were analyzed using XRD with Cu K$\alpha$ radiation ($\lambda$ = 1.5406 Å). At lower temperatures (100-200 °C), the amorphous phase appeared due to reduced atomic mobility during nucleation and film growth. In this state, atoms lack the energy required to form a regular lattice structure. Additionally, rapid cooling tends to trap atoms in a non-equilibrium state, making it impossible for the formation of a crystalline structure. These findings are consistent with reports by Yan Y. *et al.*, who observed an amorphous phase in $In_2Se_3$ deposited by RF magnetron sputtering method in the temperature range 160-250 °C[27]. Similarly, another study revealed that below 50 °C using thermal evaporation techniques followed by annealing at 373 K, 423 K, and 473 K, an amorphous phase was observed[28]. At a deposition temperature of 300 °C, a single (006) diffraction peak was observed, indicating limited crystallinity. However, upon increasing the temperature to 400 °C, eight distinct peaks corresponding to $\alpha$-$In_2Se_3$ crystallographic planes [(003), (006), (104), (009), (0012), (0015), (0018), and (0021)] emerged, signifying enhanced phase formation and improved structural ordering.

The out-of-plane lattice constant *(c)* of $\alpha$-$In_2Se_3$ was calculated from the family of 00l reflections, yielding values in the range 28.298-28.378 Å. Utilizing the 104 reflection in combination with the refined c parameter, the in-plane lattice constant *(a)* was determined to lie between 4.199 and 4.207 Å across the 300-400 °C temperature ranges. These results confirm the rhombohedral (R3m) phase of $In_2Se_3$ and indicate slight thickness-dependent variations in lattice parameters. Nyuyen T. *et al.* studied $\alpha$-$In_2Se_3$ single crystals grown by the temperature gradient method; the peaks (00l) where (l = 6, 9, 12, 15, 18, 21, 24, 27, and 30) were observed with the corresponding lattice parameters a = 4.032 Å and c = 28.705 Å[10]. Additional sources revealed that the lattice constants of $\alpha$-$In_2Se_3$ are a = 4.205 Å and c = 28.742 Å[42], while at ambient conditions, $\alpha$(3R)-$In_2Se_3$ shows a rhombohedral structure with a = 4.02377 Å and c = 28.7503 Å[11].

At the final deposition temperature (500°C), there is a transition from $\alpha$-$In_2Se_3$ to $\gamma$-InSe. All observed $\gamma$-InSe diffraction peaks appear including (0027), which was not observed at lower temperature growths as those previously identified in *Figure 2* **(a)**.

We performed X-ray fluorescence (XRF) measurements to assess the stoichiometric composition of the deposited films. As shown in *Figure 2* **(c)**, the indium-to-selenium atomic ratio varied from $0.70 \pm 0.1$ to $0.83 \pm 0.1$ as the deposition temperature increased from 100 °C to 400 °C, and reached 1.01 ± 0.1 at 500 °C. The ratios observed between 100 °C and 400 °C are in reasonable agreement with the ideal value of 0.67 for stoichiometric $In_2Se_3$, confirming near-stoichiometric growth within this temperature range. In contrast, the elevated ratio at 500 °C supports the XRD findings, which indicate a phase transition from α-$In_2Se_3$ to γ-InSe, consistent with selenium loss and indium enrichment at higher temperatures. These findings demonstrate that deposition temperature critically influences stoichiometry and crystal phase, enabling controlled transformation from α-$In_2Se_3$ to γ-InSe through thermal modulation.

The crystallite sizes were determined using the Williamson–Hall method as shown in *Figure 2* **(b)**, while *Figure 2* **(d)** illustrates the lattice parameters. Both the lattice parameters and the crystallite size exhibit variation with deposition temperature, revealing a clear temperature-dependent structural evolution in the thin films.

*Figure 3* **(a)** is the XRD pattern of γ-InSe films deposited at 500 °C with the corresponding thickness ranging from 47-564 nm on single-sided polished c-plane $Al_2O_3$ (001) substrates. As the thickness increases from 47 to 352 nm, the peak intensity grows accordingly, which is expected with film as the film thickens. However, there is a notable drop in the intensity beginning at 422 nm and continuing to 564 nm, suggesting disruption of long-range atomic order-likely which we surmise is a result of excessive thermal energy that breaks the bonds in a disordered manner.

Lattice parameter (*c*) for the deposition was derived, showing a narrow range of 28.388–28.413 Å as film thickness increased from 47 nm to 564 nm. In addition, the in-plane lattice constant (*a*) was calculated to vary between 4.199 Å and 4.203 Å, suggesting subtle lattice relaxation and thickness-dependent structural tuning [see *Figure 3* **(b)**]. Although our calculated lattice parameters differ slightly from those reported by Duan et al.[43], their study also documented variations in the in-plane lattice constant (*a*) = 4.10 Å and 4.106 Å for InSe and $In_2Se_3$, supporting the these values can shift depending on synthesis conditions and phase transitions.

The crystallite sizes [*Figure 3* **(c)**] range from 9-150 nm, with the largest size (150 nm) at 282 nm thickness and the smallest size (9 nm) at 422 nm thickness. The variation in the crystallite size values corresponds to our observations in *Figure 3* **(a)** where there is a sudden decrease in the intensities of the peaks between 422-564 nm, suggesting a structural transition.

## B.  *Tuac Plot Analysis*

We employed ultraviolet to visible (UV-vis) absorption spectroscopy to investigate the optical properties and determine the optical bandgap energy of a-$In_2Se_3$, α-$In_2Se_3$, and γ-InSe at different temperatures. To estimate the optical bandgap energy, a linear fit was applied to the fundamental absorption region, as well as to the slope below the fundamental absorption, which was used as an abscissa. The intersection point of these two fitted lines corresponds to the optical bandgap energy of the material.[36]  From *Figure 4* **(a-e)**, the direct bandgap energy of amorphous is between 2.5-2.35 eV, which is close to the reported value of 2.14 eV [44]. On the other hand, crystalline α-$In_2Se_3$ shows direct bandgaps between 1.56-1.55 eV, whereas γ-InSe has a band gap energy of 1.58 eV. This result is consistent with previously reported values of 1.453 eV [5], 1.43 eV[45] and 1.6 eV[46] for α-$In_2Se_3$, and close to 1.29 eV for γ-InSe[47], indicating variations in synthesis and measurement conditions. Additionally, other studies reported bandgap energy of bulk α-$In_2Se_3$ and γ-$In_2Se_3$ to be 1.2 eV[48,49], while few-layered InSe has an indirect band gap of 1.4 eV[49].

Further analysis of *Figure 4* **(a-e)** shows that bandgap energy is influenced by the deposition temperature. According to Qasrawi A.F., the allowed direct bandgap energy of α-$In_2Se_3$ decreases with increasing temperature, and it shifts from 1.39 eV to 1.26 eV as the temperature increases from 300 K to 480 K[50]. Our experimental results are consistent with the existing literature on the optical bandgap energy of a-In$_2$Se$_3$, α-In$_2$Se$_3$, and γ-InSe. See *Figure 4* **(f)** for the trend in bandgap evolution as a function of deposition temperature.

## C.  *Kelvin Probe Force Microscopy (KPFM) Analysis*

Kelvin probe force microscopy (KPFM) measures the contact potential difference (CPD) between a conductive probe and the sample, V$_{CPD}$, which arises from their work-function difference ΔΦ (i.e., e·V$_{CPD}$ = Φ$_{tip}$ − Φ$_{sample}$)[40,51,52]. Because Φ is defined as the energy required to move an electron from the Fermi level to vacuum, KPFM provides spatially resolved information about work function and, by extension, local surface charge, bonding, and electronic structure [53]. In air, offsets in work function from environmental factors, such as humidity can cause shifts in the report work functions, this creates a scenario where the primary uncertainty in reported values are limited in accuracy by up to ~0.5 eV offset, but spatial variation within a measurement is known to much higher precision.

*Figure 5* **(a–f)** present AFM topography, KPFM CPD maps, and representative line profiles for a-In$_2$Se$_3$. The a-In$_2$Se$_3$ films display a rounded, granular morphology consistent with the amorphous phase inferred from XRD [*Figure 2* **(a)**], with early coarsening visible as deposition temperature increases, see *Figure 5* **(a)** and *Figure 5* **(d)**. The corresponding CPD images [*Figure 5* **(b), (e)**], and line cuts [*Figure 5* **(c), (f)**] yield V$_{CPD}$ = −2.28 V and −3.55 V for the lower- and higher-temperature films, respectively. Using the calibrated tip work function (Methods), we convert these CPD values to apparent sample work functions of 5.26 eV and 6.53 eV. This range reflects the expected sensitivity of work function to local bonding, disorder, and surface termination in the amorphous state, particularly since in the 200 C⁰ films we begin to see coarsening leading to local variation in the work function, as discussed further below.

Figure 6 **(a–i)** summarize KPFM results for α-In$_2$Se$_3$ and γ- InSe. AFM reveals hexagonally faceted grains across 300–500 °C, indicative of the crystalline phase, with lateral size increasing modestly with temperature [Figure 6 **(a), (d), (g)**]. Similar hexagonal motifs have been reported for In–Se compounds grown by CVD or MOCVD and for sputtered In$_2$Se$_3$ at comparable pressures, corroborating our structural assignment.[9,54,55] The KPFM maps [Figure 6 **(b), (e), (h)**] and line profiles [Figure 6 **(c), (f), (i)**] yield V$_{CPD}$ ≈ −3.15, −3.20, and −3.65 V at 300 °C, 400 °C, and 500 °C, corresponding to apparent work functions of 6.13 and 6.18 eV for α-In$_2$Se$_3$, and 6.64 eV for γ- InSe. Research on the work function of α-In$_2$S$_3$ and γ-InSe is ongoing; hence, the experimental literature available on this topic is limited. The heterostructures of PdSe$_2$ and α-In$_2$S$_3$ show a work function difference of 0.25 eV.[56] Reporting of the work function directly of α-In$_2$S$_3$ are primarily from density functional theory, giving values of 6.13 eV[57] all the way down to 4.83 eV[53] demonstrating debate even within the theoretical community. In contrast, the α-In$_2$S$_3$ heterostructure exhibits work functions of 4.73 eV and 5.94 eV on the electrostatic potential plot[58]. Another study reported 4.80 eV as the work function of InSe.[59] Our findings correspond to the existing literature range; however, we cannot assert that our experimental findings are the utmost accurate work function of a-In$_2$Se$_3$, α-In$_2$S$_3$, and γ- InSe due to ambient conditions we expect up to a 0.5 eV variation in work function[60], but reports for more reactive surfaces up to 0.9 eV depending on oxidization state have been reported.[61] That being said, these results provide direct experimental measures of the work function of a-In$_2$Se$_3$, α-In$_2$S$_3$, and γ- InSe which are limited in the literature. We therefore interpret these results as growth parameter- and structure-dependent apparent work functions rather than definitive absolute values, recognizing known dependencies on tip calibration, surface adsorbates, and measurement mode.

Figure 7 illustrates the variation of the average contact potential and work function as a function of deposition temperature. The measured increase in work function with temperature, from 5.26 eV in the a-phase to 6.18 eV in the crystalline α-phase, and 6.64 eV in the γ-InSe indicates a shift in the electronic structure. This is due to growing crystallinity, and a more favorable band alignment at the surface. These insights deepen our understanding of the material system and give a practical way to fine-tune deposition conditions and tailor surface energetics for specific device applications.

To assess lateral homogeneity, we repeated line cuts at multiple positions (*Figure 8* and *Figure 9*). For a-In$_2$Se$_3$, repeated horizontal transects at distinct locations reproduced Φ = 5.27 eV within experimental uncertainty [*Figure 8* **(c–h)**], indicating a laterally uniform work function landscape typical of an amorphous film. However, something interesting emerges for slightly warm growth at 200 °C. In the AFM image the morphology shows some coursing of the granular structure, but the XRD remains consistent with an amorphous film. However, the work function for shifts lowers to Φ = 6.53 eV, but with clearly much lower Φ = 6.51 eV for the large grain structures that emerge. This demonstrates likely local ordering of the a-In$_2$Se$_3$ leading to slight increase in work function. For α-In$_2$Se$_3$ and γ-InSe small yet reproducible work function variations correlate with grain geometry and facet orientation. For example, at 300 °C we find 6.21 eV vs 6.22 eV for two neighboring hexagons, one lying flatter than the other [*Figure 9* **(a-d)**]. Similar, modest offsets are observed at 400 °C (6.45 and 6.30 eV), and 500 °C (6.79 and 6.7 eV) [*Figure 9* **(e-l)**]. These ~0.01-0.2 eV differences likely reflect facet-dependent surface dipoles and local carrier density/defect distributions at grain edges and vertices; importantly, they are small enough that the global work function is dominated by the phase (amorphous vs crystalline) and growth temperature, with only second-order modulation by microfacet geometry.

Overall, KPFM establishes three key points: (i) a-In$_2$Se$_3$ exhibits laterally uniform work function consistent with structural disorder; (ii) α-In$_2$Se$_3$ and γ- InSe shows weak, facet-correlated work function contrasts superimposed on a temperature-dependent morphology baseline; and (iii) the extracted apparent work functions align with literature ranges for In–Se materials while underscoring the necessity of rigorous tip calibration and environmental control when interpreting absolute Φ.

## IV. CONCLUSIONS

We employed RF magnetron sputtering to investigate the phase transition in the In-Se system, focusing on the transformation from a-In$_2$Se$_3$→ α-In$_2$Se$_3$→ γ-InSe. XRD analysis revealed that the amorphous phase forms between 100–200 °C, while the α-In$_2$Se$_3$ phase exists between 300–400 °C, and γ-InSe at 500 °C. AFM confirmed the crystallinity of the films, with hexagonal grains of α- In$_2$Se$_3$ and γ-InSe appearing in the higher temperature range. KPFM measurements showed distinct contact potential differences between phases, with work functions ranging from 5.26–6.53 eV for the a-phase, 6.13–6.18 eV for α- In$_2$Se$_3$, and 6.64 eV for γ-InSe. UV-vis spectroscopy was used to determine the bandgap energies of each phase, and XRF analysis confirmed the elemental composition. Our results provide actionable insights for refining contact engineering and advancing the integration of In$_2$Se$_3$ into next-generation phase-change and optoelectronic technologies.

## ACKNOWLEDGMENTS

This material was based upon work supported by the National Science Foundation under Grant No. 2328747.

## AUTHOR DECLARATIONS

**Conflicts of interest**

The authors have no conflicts to disclose.

**Author Contributions**

**Marius O. Eji:** Writing-Original draft (lead); Data curation (lead). **Md. Sakauat Hasan Sakib:** Data curation (supporting). **Joseph P. Corbett:** Data curation (lead); Supervision (lead); Validation (lead); Writing-review & editing (lead); funding acquisition (lead).

# DATA AVAILABILITY

The data that support the findings are available upon request from Professor Perry Corbett, Miami University.

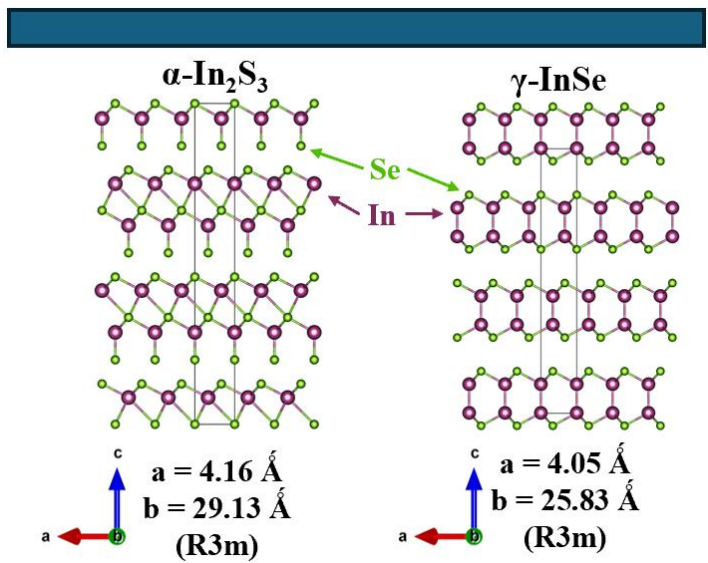

Figure 1 *3D structure of α-In$_2$Se$_3$ (left).and γ-InSe (right). The respective unit cell is shown as a black rectangle within the structure. Lattice constants and space group are provided below each structure.*

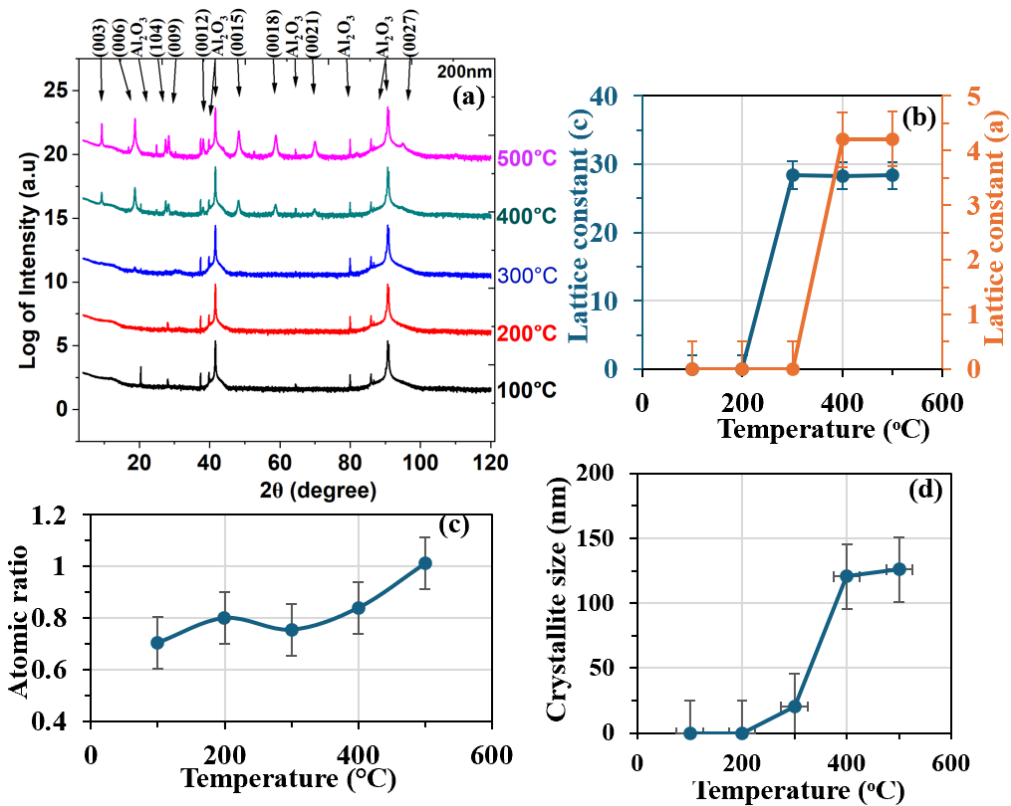

*Figure 2 presents a comprehensive structural analysis of thin films deposited on double-sided polished c-plane Al₂O₃ (001) substrates at a constant thickness of 200 nm, with varying deposition temperatures. Panel (a) shows the XRD patterns, revealing temperature-dependent changes in crystallinity and phase formation. As deposition temperature increases, the emergence and sharpening of diffraction peaks indicate enhanced crystallization and phase stability. (b) displays the calculated lattice constants a and b, which reflect subtle shifts in unit cell dimensions, suggesting temperature-induced strain relaxation. (c) shows the atomic ratio, offering insight into stoichiometric fidelity and potential deviations due to thermal effects. (d) plots the crystallite size as a function of deposition temperature, with vertical error bars denoting measurement uncertainty and solid lines representing interpolated trends. The data clearly show that atomic ratio and crystallite size increase with temperature, reinforcing the link between thermal energy and grain growth.*

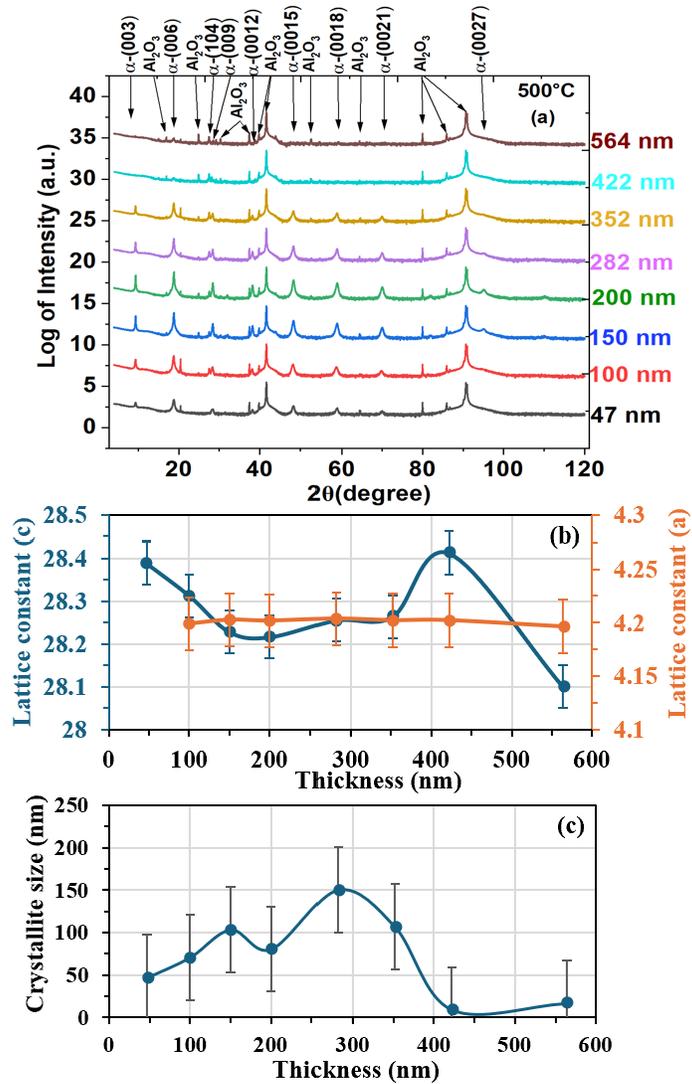

*Figure 3 Coupled symmetric scan XRD patterns, extract lattice constants and crystallite sizes. (a) XRD patterns of γ-InSe films deposited at a constant temperature of 500 °C with varying thicknesses ranging from 47 nm to 564 nm. Each curve, distinguished by color, represents one of eight samples. All films were grown on single-sided polished c-plane $Al_2O_3$ (001) substrates. (b) lattice constants as function of film thickness extract from peak positions in the 00l family to calculate c, and 104 to calculate a. (c) crystallite sizes as function of thickness computed using Williamson-Hall method from FHWM of XRD peaks.*

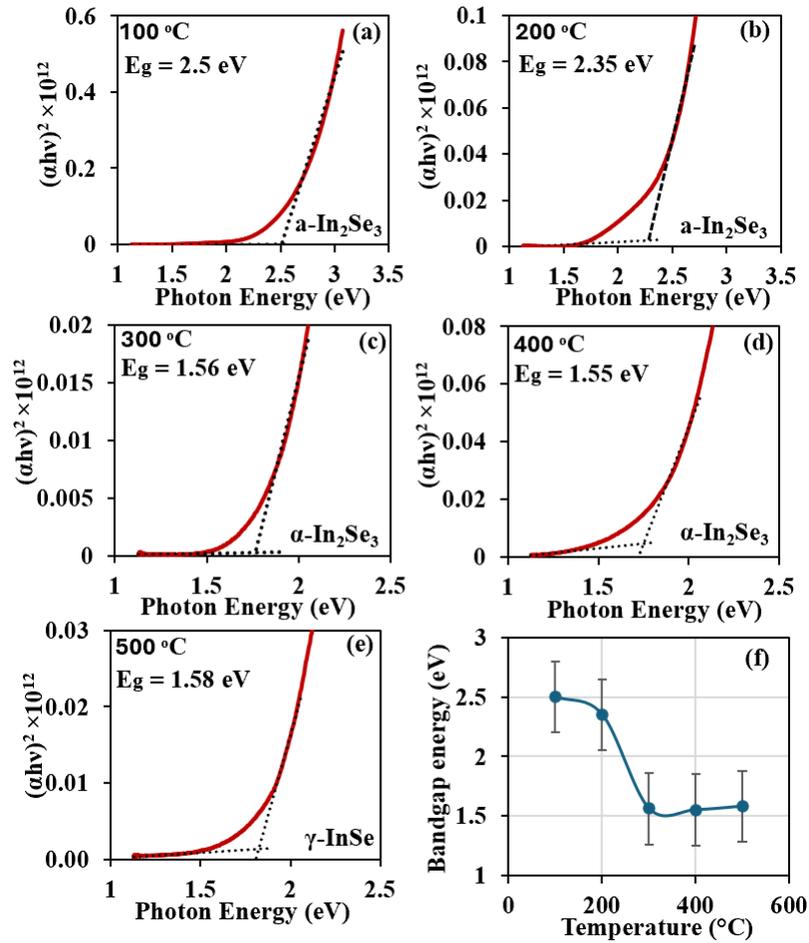

Figure 4 Tauc plots used to extract the optical bandgap energy of a- and α-$In_2Se_3$ thin films. Curves (a) and (b) correspond to amorphous phases, while curves (c–e) represent α-$In_2Se_3$. Panel (f) shows the variation of bandgap energy as a function of deposition temperature. Blue dots denote the experimental data points, the solid line represents interpolated values, and vertical bars indicate the associated error for each measurement. The results demonstrate that the bandgap energy is determined by the phase.

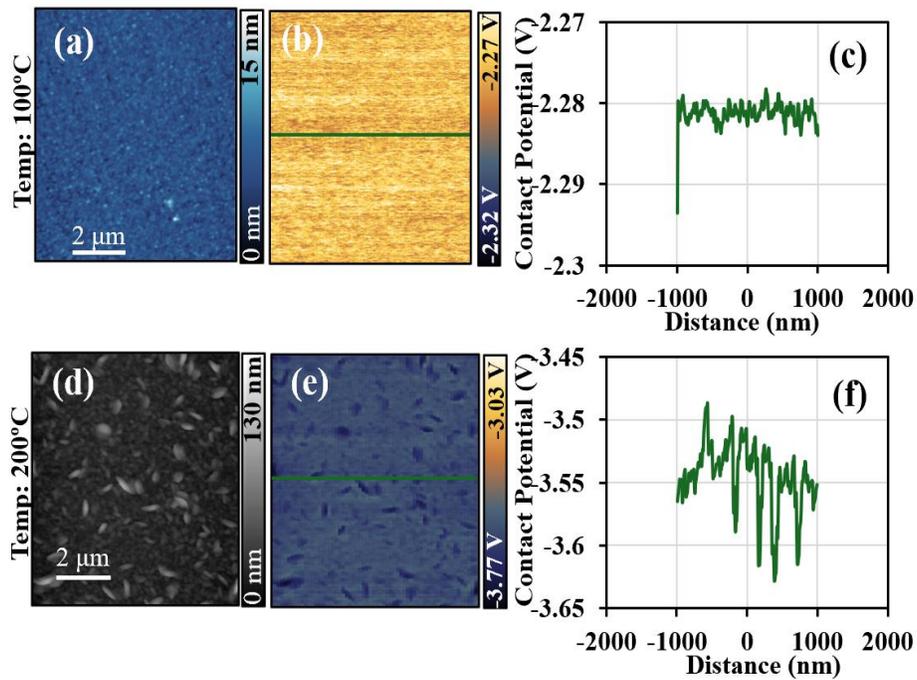

*Figure 5 illustrates the surface morphology and electronic potential landscape of a-In$_2$Se$_3$ thin films. Panels (a) and (d) show AFM images, revealing the topographical features of the films at the nanoscale. Panels (b) and (e) present KPFM contact potential maps, which visualize the spatial variation in surface potential across the same regions. (c) and (f) display line-cut plots extracted from the KPFM data, mapping correlations between surface potential and crystallographic features.*

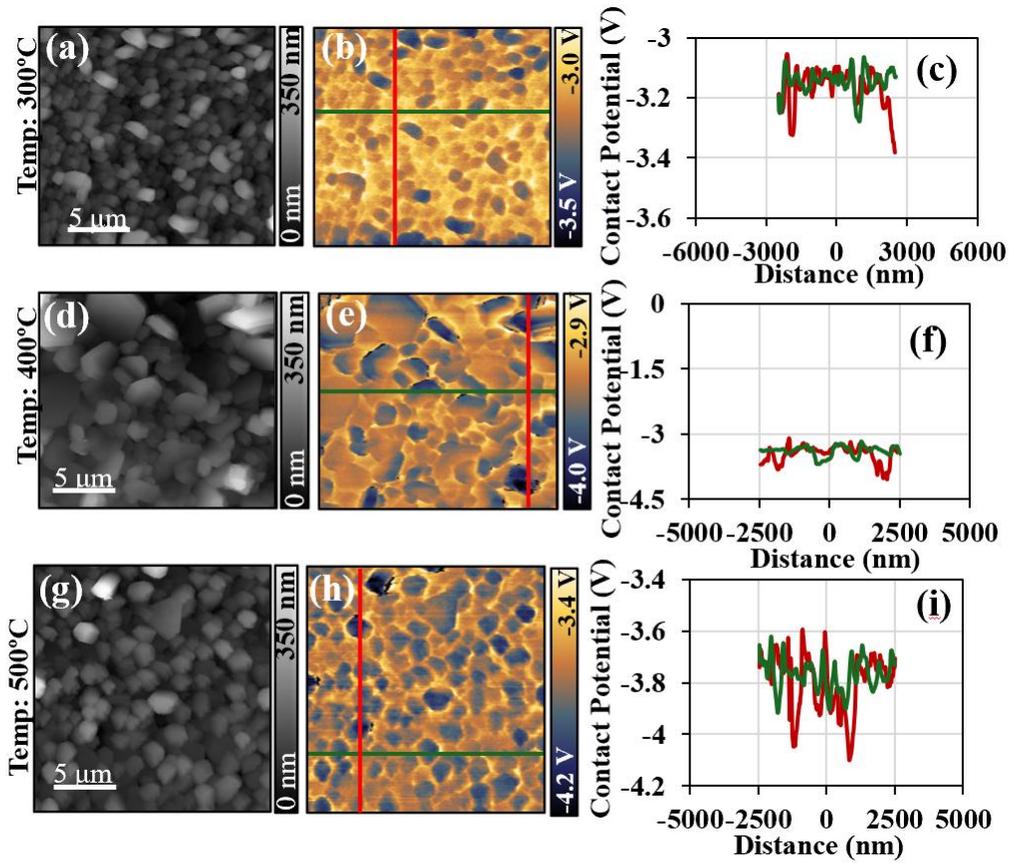

*Figure 6 presents a comparative analysis of the surface morphology and electronic potential of α-In$_2$Se$_3$ thin films, highlighting the spatial correlation between topography and contact potential. (a), (d), and (g) show AFM images, revealing the nanoscale surface features and grain structure characteristic of the crystalline α-phase. (b), (e), and (h) display KPFM maps, which visualize the distribution of surface potential across the same regions. (c), (f), and (i) provide line-cut plots extracted from the KPFM data, quantifying variations in contact potential and offering insight into electronic uniformity and phase-related contrast.*

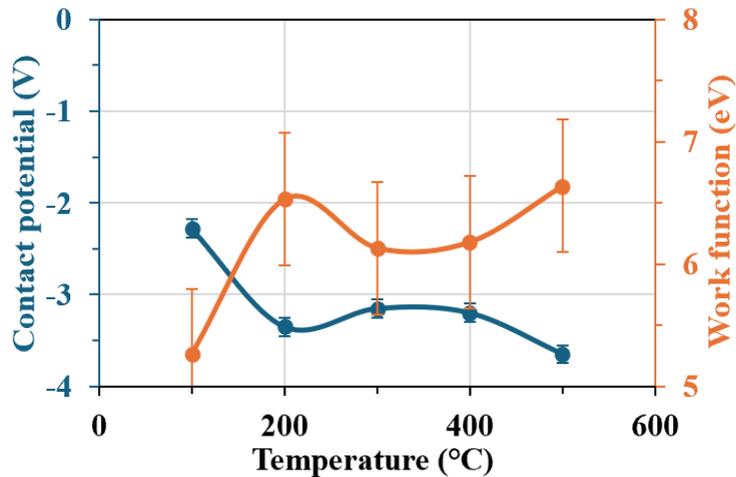

*Figure 7 illustrates the variation of the average contact potential and work function as a function of deposition temperature. The blue and orange dots represent experimentally measured data points, while the solid lines show interpolated values that trace the overall trend. Vertical error bars accompany each data point, indicating the measurement uncertainty and reinforcing the precision of the analysis.*

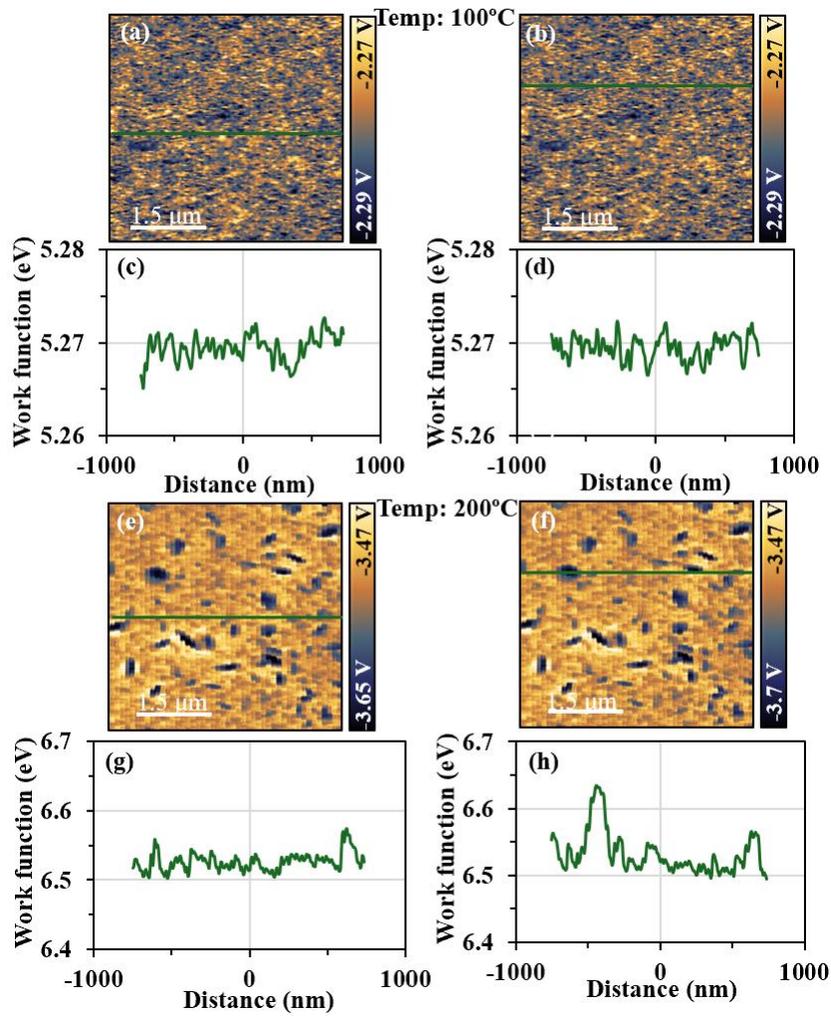

*Figure 8 presents KPFM images of a-In$_2$Se$_3$, alongside line-cut plots of the contact potential taken at multiple locations across the surface. The green lines indicate horizontal line cuts at distinct points, allowing for a comparative analysis of work function. These measurements reveal subtle fluctuations in work function, which are indicative of local electronic inhomogeneities and phase-related contrast within the amorphous structure.*

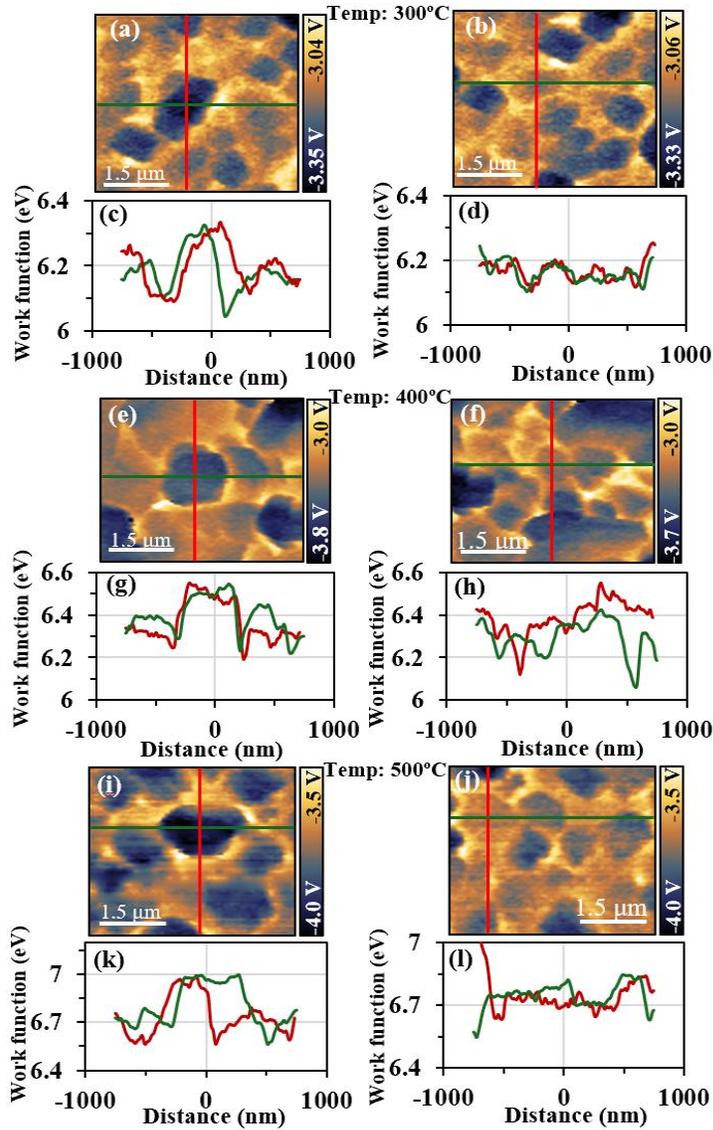

*Figure 9 presents KPFM images of α-In$_2$Se$_3$, alongside line-cut plots of the contact potential taken across two distinct hexagonal surface features. The green lines represent horizontal line cuts, while the red lines denote vertical cuts—each tracing the work function variation across different crystallographic orientations of the hexagonal domains. This figure offers a high-resolution view into the electronic heterogeneity of the α-phase, revealing how surface symmetry and grain orientation influence local work function.*